\documentclass[aps,prl,twocolumn,amsmath,amssymb,showpacs]{revtex4}
\usepackage[dvips]{graphicx}
\usepackage{ulem}
\usepackage{psfrag}
\begin{document}

\title{Breakdown of adiabaticity when loading ultra-cold
 atoms in optical lattices}
\author{Jakub Zakrzewski} 
 
\affiliation{
Instytut Fizyki im ienia Mariana Smoluchowskiego and
Mark Kac Complex Systems Research Center, 
Uniwersytet Jagiello\'nski, ulica Reymonta 4, PL-30-059 Krak\'ow, Poland}

\author{Dominique Delande}
\affiliation{Laboratoire Kastler-Brossel, UPMC, ENS, CNRS;  
4 Place Jussieu, F-75005 Paris, France}
\date{\today}


\begin{abstract}
Realistic simulations of current ultra-cold atoms experiments in optical lattices 
show that the ramping up of the optical lattice is significantly nonadiabatic, implying
that experimentally prepared Mott insulators are not really in the ground state
of the atomic system.  
The nonadiabaticity is even larger in the presence of a secondary quasi-periodic lattice simulating ``disorder''.
Alternative ramping schemes  are suggested that improve the adiabaticity when the disorder is not too large.
\end{abstract}
\maketitle

Ultra-cold atoms in optical lattices form a wonderful toolbox \cite{jaksch05} for creating novel 
matter phases realizing a ``condensed matter theorist dream''. This comes from the flexibility
of parameters in cold atom physics where both the depth of
the optical lattice (changing either the laser intensity or its detuning) 
as well as atom-atom interactions (via external magnetic 
field and Feshbach resonances) can
be modified with unprecedented precision.
By increasing the depth of the optical lattice where a Bose-Einstein condensate is 
loaded, one induces a transition between the superfluid (SF) phase and the Mott insulator (MI) phase, 
as suggested  in \cite{jaksch98} (basing on the Bose-Hubbard (BH) tight binding model \cite{fisher89})
and attempted few years later \cite{greiner02}. 
That triggered several experimental and theoretical studies involving 
bosons \cite{stoeferle04,spielman07}, fermions and their mixtures
(see \cite{lewen2007} for details). 

Disorder can be 
introduced in cold atom systems in a controllable way using optical potentials 
created by laser speckles or bichromatic lattices~\cite{damski03}.
This exciting possibility has attracted a lot of research \cite{paper-speckles}. 
For weakly (or non) interacting bosons, Anderson localization 
has been recently observed~\cite{aspect08}.
For strongly interacting bosons in a disordered lattice,  
 studies of the BH model \cite{giamarchi88,fisher89} revealed the 
existence of a novel insulating but compressible phase called Bose glass (BG) phase. 
  
As far as we know, a clear demonstration of the existence of the BG phase in traditional 
condensed matter experiments is yet to be made. In a recent experiment \cite{fallani07} using ultra-cold atoms,
a smearing of the absorption peaks was observed in the presence of ``quasi-disorder'' and interpreted as 
being consistent with the existence of a BG. While the 1D BH model in bichromatic
lattice has been discussed theoretically in detail \cite{Roscilde08}, no direct
comparison of the theory and experiment has been made. 
Such an analysis is therefore needed. 

The system can be described by the BH Hamiltonian:
\begin{equation}
\hat{H} = -J \sum_{\left<j,j'\right>}
\hat{b}_j^\dagger \hat{b}_{j'} + \frac{U}{2} \sum_{j} \hat{n}_j
\left( \hat{n}_j - 1 \right) + \sum_{j} \epsilon_j \hat{n}_j, 
\end{equation}
where $\hat{b}_j$ ($\hat{b}_j^\dagger$) is the destruction (creation) operator of one 
particle in the $j$-th site,
$\hat{n}_j=\hat{b}_j^\dagger \hat{b}_j$ is the number operator, and $\left<j,j'\right>$ 
indicates the sum on nearest neighbors
\cite{fisher89,jaksch98}.
The interaction energy $U$ and the hopping energy $J$ can be evaluated via appropriate 
integrals of Wannier functions \cite{jaksch98} obtained for lattices of a given depth. 

The sites energies, $\epsilon_j$, are given by the 
sum of  the harmonic trap potential and -- if present -- 
an additional ``quasi-disorder" due to
a second weak optical lattice created by a standing wave with wavelength $\lambda_2$:
\begin{equation}
\epsilon_j=\frac{1}{2} m \omega^2 a^2 (j-j_0)^2 + s_2 E_{R2} \sin^2 \left(\frac{\pi j\lambda_1}{\lambda_2} + \phi
\right)
\end{equation}
where $m$ is the particle mass, $a$ the lattice spacing, $\omega$ the trap frequency,
$j_0$ the position of the trap center, $E_{R2}$ the recoil energy associated 
with wavelength $\lambda_2$~\cite{fallani07}, and $\phi$ the phase of the $\lambda_2$ standing wave. 
$s_2$ measures the amplitude of the disorder in recoil energy units \cite{nota1}.

In the experiment \cite{fallani07}, an array of 1D systems is obtained by loading a $^{87}$Rb 
condensate in a 2D optical lattice (``transverse lattice''). 
The height of the 2D lattice, $s_{\perp}=40,$ results in a negligible tunneling rate between 
the one-dimensional atomic tubes. 
During the ramping on of the ``transverse lattice'',  an optical potential in 
the third direction is turned on as well, along the axes of the atomic tubes. Thus a collection 
of independent 1D atomic samples is loaded in the superposition of a ``main'' optical lattice 
(at $\lambda_1=830$~nm with a height $s_1=16$) and a ``disordered'' optical lattice 
(at $\lambda_2=1076$~nm with a height $s_2$). All lattices are  
slowly ramped up, increasing the intensity of the laser beams following an exponential ramp
over 100~ms time. 
 After the loading, the excitation spectrum of the obtained state is measured by modulating the 
intensity of the ``main'' lattice beam for 30~ms with an amplitude of 30\%. Also, the coherent fraction is
extracted from time of flight density profiles \cite{stoeferle04,spielman07,fallani07}.

Our numerical simulation of the experiment is performed using a 
TEBD algorithm~\cite{vidal03} (also known as t-DMRG algorithm \cite{tdmrg}). 
It has already been applied to the BH model~\cite{kollath06,clark06}; 
we use it here for parameters of a given experiment aiming at
reproducing the processes taking place during ramping up of the lattices potential.
Since the theory is based on the BH model, it cannot simulate the very initial stage of
the experimental procedure that starts from a harmonically trapped Bose-Einstein condensate. 
In the first stage, the atomic gas is in a SF state; ramping up the lattice 
must be done sufficiently slowly to ensure that all atoms remain in the lowest band of 
the periodic potential. The optical potential must be increased 
from 0 to one recoil energy in a time longer than the recoil  
time (300 $\mu$s), which is largely fulfilled in the experiment. 
The atomic wave function then follows 
adiabatically the shrinking of the Wannier state in each site
as long as tunneling between neigboring sites remains dominant. Following this reasoning, our simulations
are performed starting from $s_1=4$ or 5, where the
system is already in the tight binding regime, in the SF phase. We have checked that our results are
practically independent of this choice (proving that the initial stage is adiabatic).
We choose 151 particles occupying roughly 80 lattice sites, in order to match the experimental 
parameters of~\cite{fallani07}. Similar conclusions are obtained for different
values of the parameters.

In the experiment \cite{fallani07}, an absorption peak at frequency $1.9$ kHz was observed, while the interaction
energy for  $s_1=16$ is
$U=2.3$ kHz. We thus believe that the $s_1$ parameter was overestimated. 
To match the position of the peak, we modify all laser intensities by a $7/8$ 
factor yielding $s_1=14$ and $s_\perp=35.$ 
Attempts to simulate the absorption curve for the 30 \% modulation of the lattice lasting for
30 ms failed. Such a strong modulation apparently creates too much excitation and entanglement
across the lattice for the TEBD algorithm to be reliable. The absorption spectra obtained for 
10-30 times smaller modulation resemble those obtained in \cite{kollath06}
with a prominent peak around $U$ and a much smaller one around $2U$ (in experiment this relation is
reversed). 
In the presence of disorder, the absorption peaks broaden --- both in the real experiments and in
our numerical simulations --- but it is hard to extract any information from this fact:
firstly because of nonlinearities induced by the experimental strong modulation, and secondly because
absorption spectra turn out to be almost insensitive to the details of the dynamics such as a breakdown of adiabaticity.  

The numerical simulations make it possible to analyze the
state of the system by different means. 
The first step in this direction is to test an assumption done in ultra-cold atom 
`` quantum phase transition''  experiments \cite{greiner02,stoeferle04,fallani07}, 
namely the adiabaticity of the optical lattice switching on. This assumption 
is especially critical in the laser intensity region corresponding to the 
transition from the SF to the MI phase. 
For the infinite system, due to the critical slowing down~\cite{zurek}, 
such a transition cannot be fully adiabatic. For large 3D systems~\cite{greiner02}, 
an approximate mean-field simulation~\cite{zakrz05} showed that indeed adiabaticity is broken. 
For a collection of 1D tubes~\cite{stoeferle04,fallani07}, the small system size 
may assure the adiabaticity of the 
transition/crossover between phases. We can {\it directly}
test the adiabaticity by comparing the state obtained dynamically by our procedure
with the exact ground state of the system. The latter can be found by imaginary time propagation
(using essentially the same algorithm). Next, the overlap between
the dynamically created state and the ground state
can be calculated.
 
This procedure yields for a single lattice ($s_2=0$) an overlap equal to $0.095;$ for weak disorder
the corresponding overlap drops to 0.02-0.05, depending on the value of $s_2$. For $s_2$ greater than 
unity, the overlap drops dramatically to $10^{-5}$ or smaller values (note that $s_2\approx 1$ corresponds to
disappearance of the MI phase according to BH model predictions). The overlap values obtained
indicate that the switching on of the lattices is not
adiabatic. What is the character of the state prepared at the end of the ramping up? It is 
some kind of ``wavepacket", a coherent superposition of the ground state and some excited states. 
A crude way of measuring the wavepacket character is to compute the ``excess energy", i.e.
$\Delta E = E$(wavepacket)$-E$(ground state). Regardless of $s_2$ it is about 0.15$U$
meaning that the system is not highly excited.

\begin{figure}
\begin{center}
\psfrag{Overlap}{\large{$|\mathrm{Overlap}|^2$}}
\psfrag{Energy (in recoil units)}{\large{Energy (in recoil units)}}
\includegraphics[width=0.9\columnwidth]{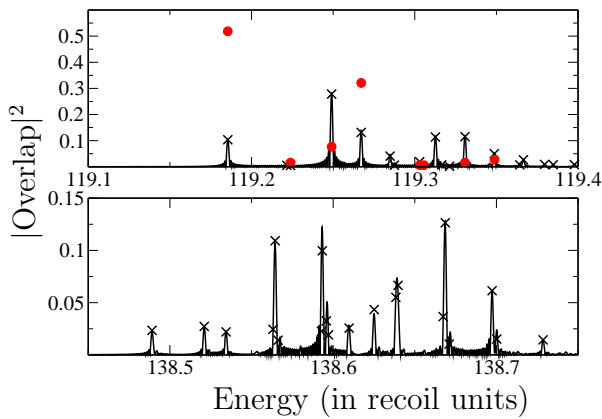}
\end{center}
\caption{(color online) Spectral analysis of the wavepacket obtained dynamically for $s_1=14$ after switching on the lattice. 
The solid line is the Fourier transform of the autocorrelation function, eq.~(\ref{eq:ct}),
showing peaks at the energy levels of the system, with an intensity
equal to the squared overlap with the wavepacket
(an exponential ramp is used). About eight (upper panel, no disorder)
or several tens (lower panel, $s_2=0.4375$) of states are significantly excited proving that the preparation is 
not adiabatic. Crosses indicate the values of overlap obtained via a 
harmonic inversion technique. Filled red circles are the corresponding overlaps for the improved ramp discussed in the
text.}
\label{hinv}
\end{figure}

More precise information can be obtained by considering the
temporal autocorrelation function of the wavepacket.
Indeed, the wavepacket can be expanded onto the eigenstates $|\varphi_i\rangle$ (with energy
$E_i$) 
of the  Hamiltonian at the
final values of $s_1$ and $s_2$:
$|\Psi\rangle = \sum_i a_i |\varphi_i\rangle.$
If we now let $|\Psi\rangle$ evolve freely with the final Hamiltonian, we obtain the
autocorrelation function:
\begin{equation}
C(t)=\langle \Psi(0)|\Psi(t)\rangle = \sum_i |a_i|^2 \exp \left( -i \frac{E_i t}{\hbar} \right).
\label{eq:ct}
\end{equation}
Thus  $E_i$'s  and overlaps can be obtained by Fourier
transforming $C(t)$. For finite time
evolution, a state of the art harmonic inversion technique
\cite{harminv} provides us with very accurate results. The exemplary results are shown in Fig.~\ref{hinv}.
In the absence of disorder, about 8 states contribute significantly to the wavepacket with
surprisingly one excited state having 30\%  overlap. For strong disorder, 
the dynamically created wavepacket spreads over several tens of eigenstates. 
There are large fluctuations of the overlaps among neigboring states,
meaning that the excitation brought by non adiabatic switching differs from a thermal excitation.
Note also that real-time TEBD propagation brings here a spectral information not accessible
to standard ``quasi-exact" numerical methods such as DMRG or Quantum Monte-Carlo.

Let us underline that, while we have analyzed as closely as possible the ramping procedure in the
recent ``disorder''
experiment \cite{fallani07}, our findings should be applicable to other similar experiments in
which both lattices and the harmonic trap are ramped on time scale of a 100 ms 
either in 1D \cite{stoeferle04} or in 3D
\cite{greiner02,white08}.

\begin{figure}
\begin{center}
\psfrag{Overlap}{\large{$|\mathrm{Overlap}|^2$}}
\psfrag{s1}{\large{$s_1$}}
\includegraphics[width=0.9\columnwidth]{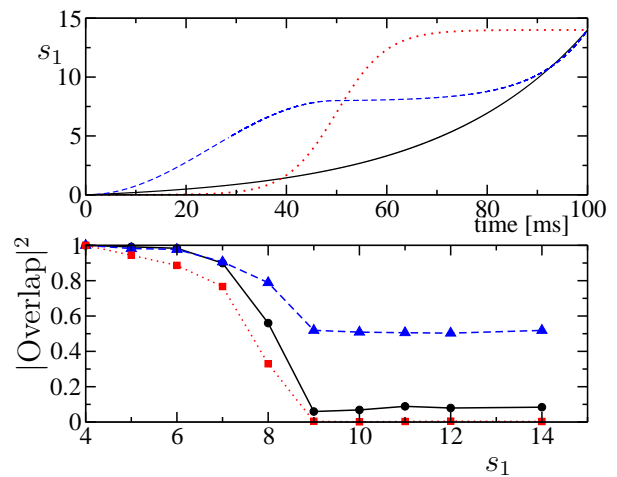}
\end{center}
\caption{(Color online)
Top panel: Different ramps used for turning on the lattices over 100 ms. The solid curve
correspond to the experimental exponential ramp with time constant $\tau=30$ ms, 
The dashed blue curve is the ``best"  pulse with a  sin$^2$ ramp up to to $s_1=8$ in 50 ms, followed by  an exponential
increase with time scale $\tau=10$ ms. The dotted red curve is the sigmoidal ramp used recently \cite{gericke07,white08}.
Bottom panel: Overlaps of the wavepacket obtained at the end of the ramp (starting from the ground state at $s_1=4$),
with ground states at different $s_1$ values. 
The evolution remains reasonably adiabatic up to $s_1=7$. The ``best" pulse performs significantly better 
than the exponential ramp yielding
more than 50\% final overlap. The sigmoidal ramp is the worst one.
}
\label{overlap}
\end{figure}

An obvious way to improve  adiabaticity is to make the ramping much slower. 
That leads, however, to other problems (e.g. decoherence due to spontaneous emission) - already a 100 ms
ramping time seems quite long. Another possibility is to optimize the ramping profile. 
The main source of nonadiabaticity is likely to come from the vicinity of the SF-MI 
phase transition. While, for a homogeneous BH model, $s_1\approx 5$ corresponds to the transition, 
the presence of the trap shifts the transition point higher in $s_1,$
see \cite{batrouni2008}). Inspection of ground states for various $s_1$
shows that the crossover point is $s_1\approx 8,$ in agreement with experimental observations \cite{fallani07}.
Therefore, a ramp which {\it slows down} around $s_1=8,$
for example  $s_1(t)=8\sin^2(2\pi t/T),$ and
again speeds up towards the final value (for example using an exponential increase) 
is likely to improve
adiabaticity. This is fully confirmed by our numerical simulations shown in Fig.~\ref{overlap}.    
The ``best" pulse shape slowing down around $s_1=8$ performs quite well,
allowing to reach  53\% overlap with the ground state,
a considerable  improvement over the exponential ramp. The excess energy for this pulse is more than twice smaller than for
the exponential pulse.
Other shapes perform almost as well provided the region around $s_1=8$ is
passed slowly. In contrast, the ``sigmoidal'' ramp of the form $s_1(t)=14/[1+\exp(-\alpha(t-t_{\rm mid})/T)]$ where 
$T=100$ms is the total ramp time, $t_{\rm mid}=T/2,$
$\alpha=20$ as in \cite{gericke07}, with a fast change in the vicinity of the phase transition,
leads to a total loss of adiabaticity. For experiments
exploring the insulator phase~\cite{white08}, this is certainly not a good choice.

While the current experiments exploring the MI regime appear not to be
strictly adiabatic, this affects slightly properties such as occupation of different lattice sites or
momenta distributions. The dynamically created wavepackets thus share with the corresponding 
ground states some global properties. In this approximate sense, the creation of MI state in 1D experiments without
disorder \cite{stoeferle04} is confirmed by our simulations.

The situation is quite different in the presence of the secondary lattice creating quasi-disorder.
Already for a relatively small disorder, $s_2=0.4375,$ a few excited states are significantly populated, 
see Fig.~\ref{hinv}. 
The situation becomes drastically worse for $s_2>1$ --- the region where MI should disappear completely. 
There, the overlap between the dynamically created wavepacket and the ground state becomes 
vanishingly small, $10^{-6}$ or less for the exponential ramp at $s_2=2.1875$. 
However, the experimental results reported in \cite{fallani07}
support the claim that a behavior characteristic of BG was observed.

\begin{figure}
\begin{center}
\psfrag{Dni}{\large{$\Delta n_i$}}
\psfrag{Avni}{\large{$\langle n_i \rangle$}}
\includegraphics[width=0.9\columnwidth]{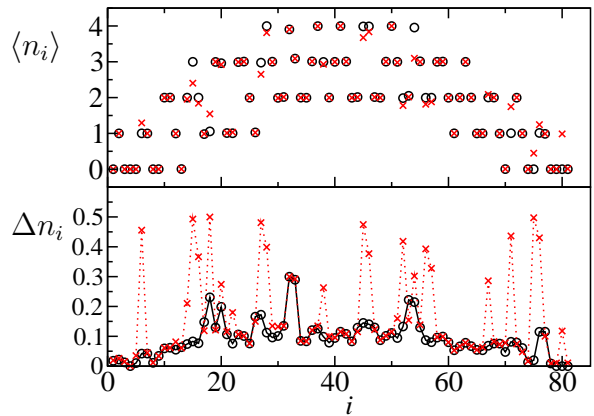}
\end{center}
\caption {(Color online) Average occupation (upper panel) and
its variance (lower panel) of individual sites, for $s_1=14$ and
$s_2=2.1875,$ i.e. for parameters where a Bose glass
phase is expected. Black circles (and solid line) correspond to the ground state
of the system, with essentially only integer occupations and small variances, together with
fluctuations of the occupations along consecutive sites, as expected for a Bose glass.
Red crosses (and dashed line) correspond to the dynamically created wavepacket. Non-integer occupations
and largely increased variances show significant excitation of the system and that its properties are
different from a Bose glass.}
\label{occup}
\end{figure}

The calculated momenta distributions averaged over different phase shifts between lattices, 
show a broad peak similar to the one for the ground state at $s_1=14$ and $s_2=2.1875$. 
The difference
between the ground state and the wavepacket becomes evident looking at the occupations of sites and
their variances $\Delta n_i=\sqrt{\langle n_i^2\rangle -\langle n_i \rangle^2}$ as shown in Fig.~\ref{occup}. 
The ground state is characterized by integer occupations with relatively low number variance, 
while the dynamically excited wavepacket
shows an average variance twice larger and non-integer occupation of some sites. The optimized
pulse, working well for no disorder, leads to similarly disappointing results.

In summary, we have shown that the experimental preparation of a MI in recent
experiment \cite{fallani07} is partially non adiabatic. Similar conclusion should apply to earlier
experiments \cite{greiner02,stoeferle04}. Still the global properties of the created wavepackets resemble
those for the ground states (in particular regions of integer occupations and low number fluctuations appear
as in the MI phase). A careful reshaping of ramps used to switch on the lattices may lead to an
adiabatic creation of MI.
 
The presence of disorder amplifies nonadiabaticity. 
Our simulations indicate that 
the properties of the ground state and of dynamically created wavepackets differ substantially indicating that a BG
was not actually achieved in the experiment \cite{fallani07}.
The observables measured in this experiment are not sufficiently sensitive
to the state of the system to characterize unambiguously a BG. To that end, other tools suggested recently \cite{rosc-we}
might prove useful.
A simple optimization of the pulse shape, successful in the absence of disorder, does not help. 
We are currently working on other schemes aiming at dynamical creation of a BG.

While competing this work we became aware of a recent work~\cite{edwards08} which shows 
that adiabaticity is much harder to obtain in the presence of two lattices 
also for shallow lattices (Gross-Pitaevskii regime).
  
We are grateful to C. Fort, L. Fallani and M. Inguscio for discussions.
DD was partially supported by IFRAF, J.Z. acknowledges support by Polish Ministry of Education and Sports (2008-2010).   
This work is realized within Marie Curie TOK scheme COCOS (MTKD-CT-2004-517186).


\begin{thebibliography}{99}
\bibitem{nota1}The heights of the optical lattice potentials $V_{\mathrm{latt}}$ will be always expressed in units 
of the laser recoil energy $E_R=h^2/(2m\lambda^2)$ through the dimensionless parameter $s=V_{\mathrm{latt}}/E_R$.
\end{thebibliography}
\end{document}